\documentclass[twocolumn,showpacs,preprintnumbers,amsmath,amssymb]{revtex4}
\usepackage{amsfonts,amsmath,amsthm,amssymb}
\usepackage[dvips]{graphicx}
\newcommand{\fspace}{\kern 1pt}
\begin{document}
\title{Randmoness and Step-like Distribution of Pile Heights
in Avalanche Models}
\author{A.~B.~Shapoval}
 \altaffiliation{Finance Academy under the Government of the Russian Federation.}
 \email{shapoval@mccme.ru}
\author{M.~G.~Shnirman}
 \altaffiliation{Institut de Physique du Globe de Paris}
 \email{shnir@mitp.ru, shnirman@ipgp.jussieu.fr}
\affiliation{International Institute of Earthquake Prediction Theory
and Mathematical Geophysics, Warshavskoye sh.~79, kor.~2, Moscow, 117556, Russia.}

\date{\today}


\begin{abstract}
The paper develops one-parametric family of the sand-piles
dealing with the grains' local losses on the \textit{fixed amount}.
The family exhibits the crossover
between the models with deterministic and stochastic relaxation.
The mean height of the pile is destined to describe the crossover.
The height's densities corresponding to the models with relaxation
of the both types tend one to another as the parameter increases.
The densities follow a step-like behaviour in contrast to the peaked shape
found in the models with the local loss of the grains down
to the \textit{fixed level}
[S.~L{\"u}beck, Phys. Rev.~E, 62, (2000), 6149].
A spectral approach based on the long-run properties of the pile height
considers the models with deterministic and random relaxation
more accurately and
distinguishes the both cases up to admissible parameter values.
\end{abstract}

\maketitle
\section{Introduction}
In 1987 Bak et al (BTW) introduced their sand-pile model \cite{BTW87}.
The model determines a system containing some physical quantity
called sand in the original paper.
The system is slowly loaded. Extra loading results in a local relaxation.
The local relaxation releases energy that
can instantly spread out to the large distances. The spreading
mechanism is fully deterministic.
The model's system achieves its critical state without tuning any parameter
\cite{BTW88}.

Numerous power laws describe the critical state.
They have been established theoretically \cite{D90,D99}
and numerically \cite{GM90,BMM01,KLGP00}.
The model laws find their application for such different fields
as neural networks \cite{MH91}, earthquakes \cite{BTW87G},
and solar flares \cite{BS03}.

A great demand for the model has resulted in its modifications.
The closest versions to the original sand-pile are, probably,
Manna's and Zhang's models \cite{M90,Zh89}.
Manna has defined the spreading of the local relaxation in
a stochastic way.
Zhang has introduced a continuous sand-pile. The critical behaviour of
these models exhibits a certain similarity.
Since minor changes in the model rules weakly influence the critical
behaviour \cite{B-HB96} and the number of the changes is inexhaustible,
the models need a strict classification.

Many papers assign arbitrary two models to the same universality class
if they have the same set of the exponents determining the
critical power laws \cite{B-HB96}.
The ''heat'' discussion \cite{PVZ94,ChSVZ99,IPVZ99,TMS99,Lu001}
based on the different approaches has turned
to the conclusion that BTW's and Manna's sand-piles
belong to the different universality classes \cite{ChSVZ99}.
Preliminary investigation suggests that BTW's and Zhang's
sand-piles should represent the same universality class \cite{Lu97}.

The paper \cite{Lu00} has introduced the family of the models realizing
the crossover between Manna's and Zhang's sand-piles.
The control parameter deals with the energy of the local relaxation.
The small values of the energy characterized Manna model while the
extremely big values lead to Zhang-type behaviour.
The properties of the sand distribution over the system reflect
the crossover.

The local relaxation is defined for \cite{Lu00}'sand-piles
as the loss of the energy down to the \textit{fixed level}.
The family of the sand-piles in \cite{ShSh0502} deals with
the loss of energy \textit{on the fixed amount}.

According to \cite{ShSh0502}, its family of the models makes
the crossover between BTW's sand-pile and the random walk
crossing some modification of Manna's sand-pile.
This crossover corresponds to the relatively small energy mentioned above.
On the other hand, the family determines a sophisticated limit behaviour
as the energy tends to infinity.
The classification based on the power laws fails to describe
a great diversity appeared in this continuous sand-pile family.
The paper \cite{RZh99} has introduced an appropriate global
functional, whose evolution calculated in terms of the spectrum
determines the system dynamics.

The paper \cite{KMS05} introduces the energy propagation mechanism
that lacks the local symmetry, thus leading to the sand-piles
with the quenched disorder.
These models depend on some parameter exhibiting
the value of the asymmetry. BTW sand-pile belongs to the
family of the models and corresponds to the absense of the asymmetry.
Establishing the deterministic relaxation the \cite{KMS05}'s
family of the models principally differs from that of
\cite{Lu00} and \cite{ShSh0502}.
The properties of this family are found
but the models' crossover to BTW sand-pile is not considered.

In this paper we investigate the sand-pile family of \cite{ShSh0502}.
The family's sand distribution is found to be principally different
from that appearing in the usually discussed models,
in particular, in Zhang's and Ref.~\cite{Lu00}'s models.
As the control parameter is relatively big the difference
between the deterministic and stochastic relaxation almost
disappears. Then the sand distribution over the system
is proved to become similar for the both types of the models.
However the spectral properties select the sand-piles with
the deterministic relaxation.

\section{Model}
The model deals with a two-dimensional square lattice $L\times L$.
Each cell contains $h_{ij}$ grains, where $h_{ij}$ is
less than an integer threshold $H$. At each time moment a cell $(i,j)$
is chosen at random. Its number of the grains (further, height)
$h_{ij}$ increases on $1$:
$$
 h_{ij}\longrightarrow h_{ij}+1.
 \label{e:fall}
$$
If the resulting height $h_{ij}$ remains less than $H$, then
nothing more happens at the moment. Otherwise, the cell $(i,j)$
becomes unstable and relaxes. Relaxation depends on an integer control
parameter $n$.
The unstable cell distributes its $n$ grains ``in equal parts''
among its $4$ nearest neighbours. Namely, as $n=4k$ each neighbour
gets exactly $k$ grains.

Naturally, there exist the numbers $n=4k+r$, where the residue
$r<4$ is not equal to zero. Then each neighbour gets $k$ grains
and the rest $r$ grains are distributed to four different
neighbours at random.
The following formula expresses the idea of the construction:
\begin{gather*}
 h_{ij}\longrightarrow h_{ij}-n,\\
 h_{\hbox{\scriptsize neighbour$(i,j)$}}
 \longrightarrow h_{\hbox{\scriptsize neighbour$(i,j)$}}+k
 \quad\hbox{or}
 \\
 h_{\hbox{\scriptsize neighbour$(i,j)$}}
 \longrightarrow h_{\hbox{\scriptsize neighbour$(i,j)$}}+k+1.
\end{gather*}

During this relaxation other cells can achieve the threshold $H$ and
become unstable. Then they relax according to the same rules.
If a boundary cell relaxes, $[n/4]$ or $[n/4]+1$ grains leave
the lattice and dissipate, where $[x]$ is the integer part of $x$.
(The dissipation is bigger for the corner cells).

The sequential acts of relaxation are ca\-l\-l\-ed an
\textit{av\-a\-l\-a\-n\-che}.
The \textit{size} of any avalanche is the number of the unstable
cells during the avalanche counting according to their multiplicity.
The dissipation at the boundary assures that the avalanches are defined
correctly and their size is finite.

The case of $n=H=4$ corresponds to the original sand-pile of \cite{BTW87}.
It worth noting that the paper \cite{Lu00} introduces relaxation with
the loss of any unstable cell down to the fixed level.
Namely, $h_{ij}\longrightarrow H-n$ with this $h_{ij}$ grains passed
one-by-one to the nearest neighbours; the receiver for each grain
is found at random.
These changes in the rules qualitatively influence the systemThe limit behaviour ($n\rightarrow\infty$) of \cite{Lu00}'s
family and the models in question is different.

The family really depends only on parameter $n$.
The values of $H$ does not influence the system dynamics.
They determine the admissible interval $[H-n,H)$ of the heights.

\section{Pile Height Densities}
\begin{figure}
\hbox to \hsize{\hfil
\includegraphics[height=60mm]{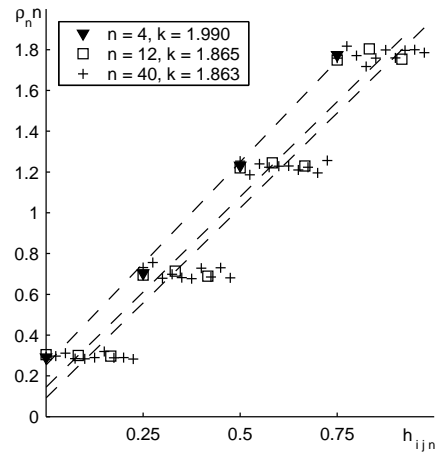}
\hfil}
\caption{\label{f:hsample4}
 Normalized density of the heights; the dashed lines are the linear fits,
 $L=256$.
}
\end{figure}

Following the ideas of \cite{Lu00} we establish some features
of the sand distribution over the lattice.
Let the normalized heights  $h_{ijn}$ be
$h_{ijn}=(h_{ij}-H+n)/n$. Then $h_{ijn}\in[0,1)$.  Let a function
$\rho_n(\cdot)$ be the density of the normalized heights $h_{ijn}$:
$\rho_n(k/n)$ is the number of $h_{ijn}$
being equal to $k/n$.
Then the normalized densities $\rho_n(\cdot) n$ follow four flat steps
with a high accuracy (fig.~\ref{f:hsample4}) for $n=4k$. The steps correspond
to the values of the density $\rho_4$ for BTW's sand-pile.
(The points of $\rho_4$ shown in fig.~\ref{f:hsample4} are in good
agreement with the analytical values found in \cite{P94}).

Each density is fitted by a linear function (fig.~\ref{f:hsample4}).
The slopes slightly increase as $n$ goes up and must be saturating
as $n$ tends to infinity.

The following construction manages to compare quantitatively four steps
of $\rho_n(\cdot)$ as $n>4$ with the four values of $\rho_4(\cdot)$
representing BTW sand-pile. Given $n=4,5,\ldots$,
the values of the density $\rho_n(\cdot)$ is sampled into four bins
$[0,0.25)$, $[0.25,0.5)$, $[0.5,0.75)$, and $[0.75,1)$ with reported
values $\psi_n(\cdot)$ at $0$, $0.25$, $0.5$, and $0.75$ respectively.
For example, $\psi_n(0)=\rho_n(0)+\rho_n(1/n)+\ldots+\rho_n(j_0/n)$,
where $j_0$ is the biggest integer with $j_0/n<0.25$.
In particular, $\psi_4(\cdot)$ coincides with $\rho_4(\cdot)$.
Then $\psi_n(\cdot)$ is defined in $4$ points. Each value
$\psi(k/4)$, $k=0,1,2,3$, represents one step of $\rho_n(\cdot)$.

Let
$$
 \sigma_n=\left.\left(\sqrt{\sum_{k=0}^3\left(\psi_n(k/4)-\rho_4(k/4)\right)^2}
 \right)\,\right/3.
$$
Then $\sigma_n$ measures the difference between the steps and
the function $\rho_4(\cdot)$ corresponding to BTW sandpile.

\begin{figure}
\hbox to \hsize{\hfil
 \includegraphics[height=60mm]{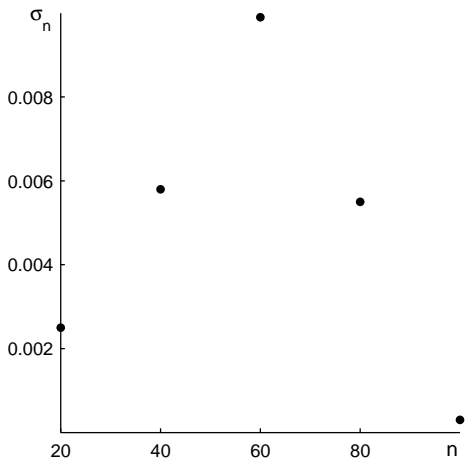}
\hfil}
\caption{\label{f:nsigma}
 Difference between the steps of $\rho_n$ and the function
 $\rho_4$ measured by the functional $\sigma_n$.
}
\end{figure}

The difference from BTW sand-pile increases as $\sigma_n$ goes up
(fig.~\ref{f:nsigma}).
The reported values of $\sigma_n$ indicate that
there exist a certain maximum of the difference corresponding
to $n\approx 60$.
This observation agrees with the change of the tendency exhibited
by the sand-pile family of \cite{Lu00} for the intermediate values of the
control parameter.

\begin{figure}
\hbox to \hsize{\hfil
\includegraphics[height=60mm]{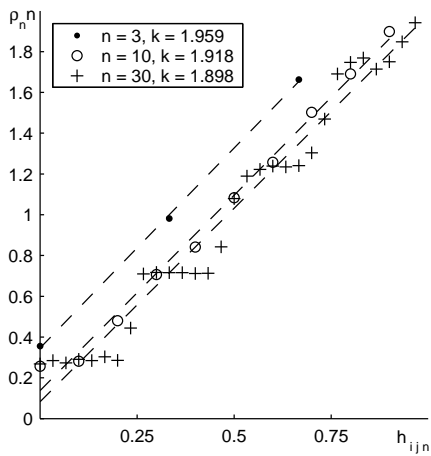}
\hfil}
\caption{\label{f:hsample}
 Normalized density of the heights; the dashed lines are the linear fits,
 $L=256$.
}
\end{figure}

In the same way, the normalized densities are introduced to develop
the models with $n\ne 4k$ starting with $n=3$.
The computer experiment proves (fig.~\ref{f:hsample})
that the densities $\rho_n$
quickly (as $n$ goes up) achieves four steps of $\rho_4$.
The linear fits have slopes appearing close to that for $\rho_{4k}$.

The following reasoning gives a rough explanation of the step-like behaviour
of the densities. Given the parameter value $n$, the heights are naturally
divided into four intervals $[H-n, H-3n/4)$, $[H-3n/4, H-n/2)$,
$[H-n/2, H-n/4)$, $[H-n/4, H)$ since each act of relaxation
maps one interval into another (possibly, excepting the boundary values
due to the random effect). The normalization (on $n$) transforms these
intervals to the domain of definition of four steps in fig.~\ref{f:hsample4},
and \ref{f:hsample}.

It worth noting that the sand-piles of \cite{Zh89,Lu00} demonstrate peaked
densities of the average height in contrast to our $\rho_n$.

So, in terms of the sand distribution the model family exhibits
a certain similarity. There exist possibilities to select the
limit behaviour ($n\to\infty$) and define an intermediate deviation
from the extreme cases.
The densities $\rho_n$ for $n=4k$ and $n\ne 4k$
become almost indistinguishable as $n$ has the order of dozens.

\section{Spectral approach}
Another approach gives evidence of the diversity of
the two cases ($n=4k$ and $n\ne4k$).  It deals with the
spectrum of the average height $h=L^{-2}\sum_{i,j=1}^L h_{ij}$.
The average height is calculated at the end of each
time moment and treated as the function on time, $h(t)$.
The paper \cite{RZh99} has successfully used $h(t)$'s spectrum
to describe the paper's sand-pile's dynamics.
Our $h(t)$'s spectrum appears to be noisy, therefore it is
averaged over its several realizations.
Besides, each $h(t)$'s realization is stored in the bins of some
length $\Delta$. Then the Fourier transform determines the spectrum $\xi$.
The highest frequency is $1/\Delta$ in the case.

A formal procedure applied for the spectrum calculation consists
of four steps.
\begin{enumerate}
\item For some fixed $N$ the time moments, which $h(t)$ is
catalogued at, is divided into $N$ non-intersecting intervals
of the same length $T$.
\item Given any fixed interval of step~$1$, the averaging of
$h(t)$-values over relatively small sub-intervals is applied to reduce
the number of data for further numerical application of
the fast Fourier transform. Let $\Delta$ be the length of the small
sub-intervals and $r$ be their quantity, $r\Delta =T$.
Then a signal $x_j$ is defined as the arithmetic mean of all
the $h(t)$-values in the $j$-th small sub-interval, $j=1,2,\ldots,r$.
\item The Fourier transform determines the spectrum of $x_j$. Let
$$
\hat{x}_k=\frac{1}{\sqrt{r}}\sum_{j=1}^r\big(x_j-\langle x\rangle\big)
e^{-\frac{2\pi ikj}{r}},
$$
where $\langle x\rangle$ is the mean of $x_j$, $j=1,\ldots,r$, $i=\sqrt{-1}$.
Then the frequencies $f_k$ are $k/(r\Delta)$, $k=1,2,\ldots,r$,
and the power spectrum $\xi(f_k)$ of the signal $x_j$
is defined as $\xi(f_k)=|\hat{x}_k|^2$.
\item Each of $N$ intervals defined in step~$1$ generates its own spectrum
$\xi(f_k)$. The averaging of $\xi(f_k)$ results in the stabilized spectrum,
which the construction aims at. The stabilized spectrum depends on
the model parameter $n$. Thus, the notation $\xi_n(f_k)$ is keeping further
for the stabilized spectrum.
\end{enumerate}

\begin{figure}
\hbox to \hsize{\hfil
\includegraphics[height=60mm]{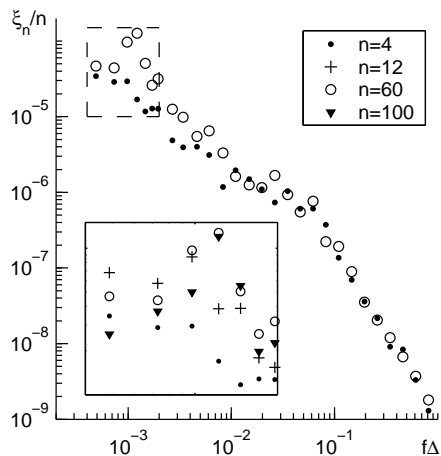}
\hfil}
\caption{\label{f:spectr}
 Normalized spectrum $\xi_n/n$ vs normalized frequencies $f=f_k$.
 The inset contains the graphs in the boxed part of the figure;
 $\Delta=8n$, $N=16$, $r=8192$, $L=256$.
}
\end{figure}

According to \cite{ShSh0502}, the family of the spectra admits a certain
normalization. If $\Delta$ is proportional to $n$, the exhibition
of $\xi_n(f_k)/n$ versus dimensionless frequencies $f_k \Delta$
collapses the graphs on the interval of the high frequencies and remains
reasonable on the other interval (fig.~\ref{f:spectr}).

Leaving the main part of the spectrum as it is we stress attention
on some interval of the low frequencies.
The corresponding part of the graphs is boxed in fig.~\ref{f:spectr} and put in its inset.
In this part the graphs (for $n=4k$) establish a noticeable rise as $n$ goes up
to the value being close to $60$. Then the tendency principally changes.
The graphs turn down to the spectrum $\xi_4$ at their left part
while demonstrating a definite peak (fig.~\ref{f:spectr}, inset).
It agrees with the behaviour of $\sigma_n$ reported earlier.

The spectrum's propagation to the left needs much bigger
$h(t)$'s domain of definition that is hardly possible.
On the other hand, the highest achieved frequencies correspond to
the avalanches of the biggest size observed during computer simulation.
Bigger avalanches are practically not observable
even for much longer simulation.
Then the computer experiment must be producing the constant spectrum
just at the left of the obtained graphs' part in fig.~\ref{f:spectr}.

The horizontal coordinates of the investigated box are not absolute
constants. They depend on the lattice length $L$ and the simulation time.
As $L$ increases the box moves to the left becoming invisible in fig.~\ref{f:spectr}.

The influence of the simulation time is essential.
The investigated part of the spectrum reflects the frequencies
of the rare, big, and strongly dissipative avalanches.
The lack of the data limits  the results covering these avalanches.
It concerns the partial conclusions about the rare avalanches in
BTW's and Manna's sand-pile \cite{MST98}.

\begin{figure*}
\hbox to \hsize{\hfil
\includegraphics[height=60mm]{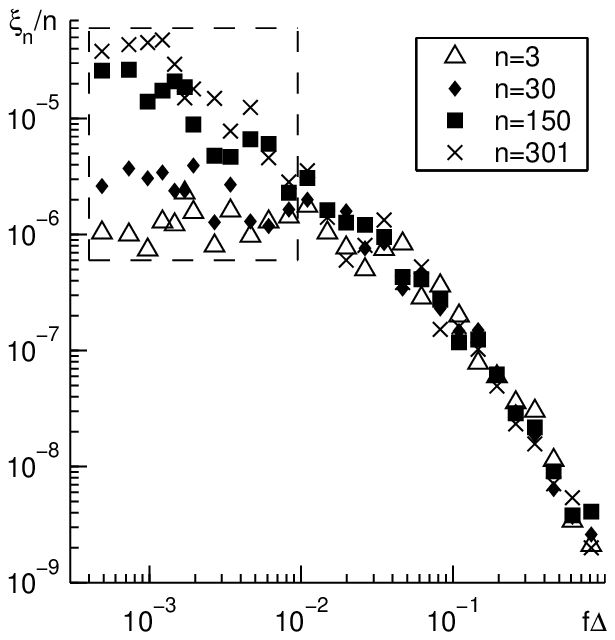}
\hfil%
\includegraphics[height=60mm]{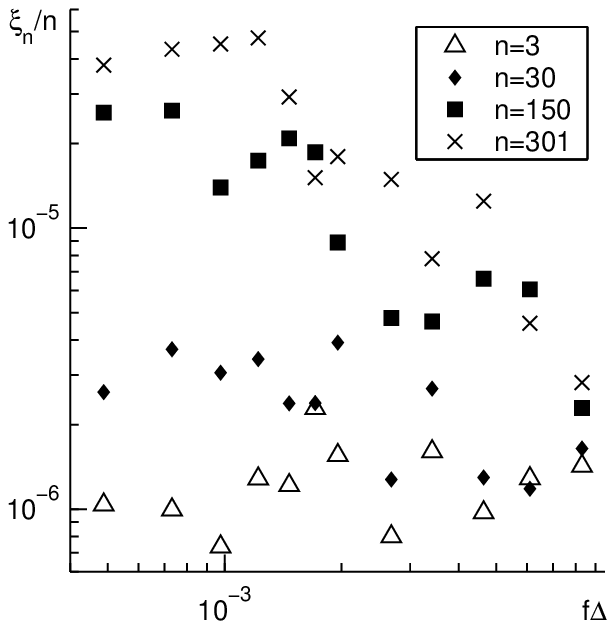}
\hfil%
}
\caption{\label{f:spectrh}
 Normalized spectrum $\xi_n/n$ vs normalized frequencies $f=f_k$;
 $\Delta=8n$, $N_R=16$, $r=8192$, $L=256$ (a) full graph,
 (b) box of fig.~a.
}
\end{figure*}

In contrast to the quick convergence of $\rho_n n$, $n\ne 4k$,
to their four values, the family $\xi_n/n$ of the spectra
exhibits quite a different tendency. Since the high-frequency spectra
are rather similar (fig.~\ref{f:spectrh}\fspace a),
the analysis is focused on the low frequencies (fig.~\ref{f:spectrh}\fspace b)
corresponding to the boxed part in fig.~\ref{f:spectrh}\fspace a.

As $n=3$ the normalized spectrum has its horizontal interval.
While $n$ increasing, $\xi_n/n$ changes on this interval.
Fig.~\ref{f:spectrh} demonstrates the minor changes as $n=30$
and completely another behaviour as $n$ is equal to $150$ and $301$.
So, for the simulated big $n$s the low-frequency spectrum essentially
deviates from $\xi_3$ as well as from $\xi_{4k}$ for big $k$.

According to fig.~\ref{f:spectr} and~\ref{f:spectrh} the long-time
evolution of the sand-pile exhibits a great complexity.
In contrast to the space features,
several patterns do not exhaust the spectrum behaviour.
Up to the developed experiments the sand-piles with the deterministic
and stochastic relaxation have the different spectra.

\section{Conclusion}
Summarizing, we develop the family of the sand-piles.
The control parameter $n$ is the number of the grains that any unstable
cell passes to its neighbours.
As $n=4k$, the propagation of the grains through the lattice
is fully deterministic despite the models with $n\ne 4k$ involve some random
effect. In terms of the sand distribution the random effect
disappears as $n$ is sufficiently big.
However the trace of the deterministic relaxation remains visible
in terms of the average height's spectrum, selecting some peak
at low-frequency spectrum.
With their peculiar evolutionary properties the sand-piles corresponding
to the deterministic relaxation may admit a certain prediction.
This hypothesis agrees with the effective precursors
found for BTW's sand-pile in \cite{ShSh04}.

\bibliography{text}
\end{document}